\documentclass[amsmath,amssymb,nofootinbib]{revtex4}
\pdfoutput=1
\usepackage{graphicx}
\usepackage{epsfig}
\usepackage{rotate}
\usepackage{amsmath}
\usepackage{amssymb}
\usepackage{amsfonts}
\usepackage{enumerate}
\usepackage{afterpage}
\usepackage{color}
\usepackage{amscd}

\newcommand{\bx}{{\bf{x}}}
\newcommand{\bq}{{\bf{q}}}
\newcommand{\bk}{{\bf{k}}}

\newcommand{\ud}{\mathrm{d}}
\newcommand{\p}{\partial}
\newcommand{\cH}{\mathcal{H}}

\def\be{\begin{equation}}
\def\ee{\end{equation}}
\def\bea{\begin{eqnarray}}
\def\eea{\end{eqnarray}}

\begin{document}

\title{Galaxy bias and gauges at second order  in General Relativity}

\author{Daniele Bertacca$^{a,b}$, Nicola Bartolo$^{c, d}$, Marco Bruni$^{e}$, \\
Kazuya Koyama$^{e}$, Roy Maartens$^{a,e}$, Sabino Matarrese$^{c,d,f}$, Misao Sasaki$^{g}$, David Wands$^{e}$,\\~}

\affiliation{
$^a$Physics Department, University of the Western Cape, Cape Town 7535, South Africa\\
$^b$Argelander-Institut f\"ur Astronomie, Auf dem H\"ugel 71, D-53121 Bonn, Germany\\
$^c$Dipartimento di Fisica Galileo Galilei, Universit\`{a} di Padova,  I-35131 Padova, Italy\\
$^d$INFN Sezione di Padova,  I-35131 Padova, Italy\\
$^e$Institute of Cosmology \& Gravitation, University of Portsmouth, Portsmouth PO1 3FX, UK\\
$^f$Gran Sasso Science Institute, INFN, I-67100 L'Aquila, Italy\\
$^g$Yukawa Institute for Theoretical Physics, Kyoto University, Kyoto 606-8502, Japan}

\begin{abstract}

We discuss the question of gauge choice when analysing  relativistic density perturbations at second order. We compare Newtonian and General Relativistic approaches. Some misconceptions in the recent literature are addressed. We show that  the comoving-synchronous gauge  is the unique gauge in General Relativity that corresponds to the Lagrangian frame and is entirely appropriate to describe the matter overdensity at second order. The comoving-synchronous gauge is the simplest gauge in which to describe Lagrangian bias at second order.

\end{abstract}

\date{\today}
\begin{flushright}
{YITP-15-2}
\end{flushright}

\maketitle

\section{INTRODUCTION}

Future galaxy surveys will probe scales comparable to the Hubble horizon, and it is therefore important to correctly incorporate relativistic effects in galaxy number counts. There are two types of relativistic corrections to Newtonian approximations. 

Firstly, relativistic corrections arise from nonlinear constraint equations in General Relativity (GR)
\cite{Bartolo:2005xa, Verde:2009hy, Bruni:2013qta, Bruni:2014xma,Villa:2014foa} (see also \cite{Fitzpatrick:2009ci,Uggla:2013kya} for related results). In a Newtonian analysis, the Poisson equation is a linear relation between the gravitational potential and the matter overdensity. In GR, this is replaced by a nonlinear relation between the metric perturbation and the matter overdensity. 
Due to this nonlinear relation, the density field becomes non-Gaussian even if the primordial curvature perturbation is Gaussian.
The non-Gaussianity in the comoving-synchronous density field can be parametried by the effective local parameter \cite{Bartolo:2005xa, Verde:2009hy, Bruni:2013qta, Bruni:2014xma,Villa:2014foa}
\be\label{fgr}
f_{\rm NL}^{\rm GR}=-{5\over3} \,.
\ee

The second type of GR correction arises because we observe galaxies on the past lightcone and not a constant time hypersurface. On Hubble scales, the number overdensity is gauge-dependent, and none of the standard gauge-invariant choices corresponds to the actual observed overdensity, which is automatically gauge-invariant. This observed overdensity
includes all redshift, lensing and volume distortions, some of which make a growing contribution on large scales \cite{Yoo:2010ni, Bonvin:2011bg, Challinor:2011bk}. The first-order GR lightcone effects contaminate the signal from primordial non-Gaussianity and so these effects must be included in order to extract the primordial signal
\cite{Bruni:2011ta,Jeong:2011as,Yoo:2012se,Raccanelli:2013dza, Camera:2014bwa,Camera:2014sba}. 

Recently, several papers have extended the calculation of the observed galaxy number overdensity on cosmological scales up to second order in relativistic perturbation theory \cite{Bertacca:2014dra, Bertacca:2014wga, Yoo:2014sfa, DiDio:2014lka, Bertacca:2014hwa}.
These results could be important to make accurate estimates of cosmological parameters, including primordial non-Gaussianity, from large-scale structure. 

Since the observed galaxy overdensity does not depend on gauge, we can calculate it using any gauge. However, a subtlety arises when we define the galaxy bias. Local bias should be defined in the rest-frame of cold dark matter (CDM), which is assumed to coincide with the rest frame of galaxies on large scales. Local bias can be computed using the peak-background split approach, where halo collapse occurs when small-scale peaks in the density exceed a critical value \cite{Press:1973iz}. Long-wavelength modes modulate this critical value. The spherical collapse model has an exact GR interpretation and the criterion for collapse of a local overdensity is when the linearly evolved density in the comoving-synchronous gauge reaches a critical value \cite{Matarrese:1997ay, Bartolo:2003gh, Wands:2009ex, Bartolo:2010rw}.  Thus it is natural to define bias in terms of the density in the comoving-synchronous gauge \cite{Challinor:2011bk,Bruni:2011ta,Jeong:2011as}. At first order, the galaxy overdensity in comoving-synchronous gauge is given on large scales and for Gaussian primordial fluctuations by
\be
\delta_{g\,C}(z,{\bf x})=b(z)\delta_C(z,{\bf x}),
\ee
where $\delta_C$ is the matter overdensity. 
This approach was extended to second order  in \cite{Bertacca:2014dra}. 

Recently, the choice of the comoving-synchronous gauge at second order has been criticised \cite{Yoo:2014vta, Biern:2014zja, Hwang:2014qfa}, 
with claims that it leads to violation of mass conservation and is inappropriate for defining galaxy bias.
Instead,  \cite{Yoo:2014vta} advocated that the correct gauge to choose is the total matter gauge \cite{Hwang:2006iw,Biern:2014zja, Hwang:2014qfa}.  
Here we show that, on the contrary,
\begin{itemize}
\item
there is {\em no} violation of mass conservation in comoving-synchronous gauge, \item
the comoving-synchronous gauge is the appropriate gauge for defining the local Lagrangian bias at second order.
 \end{itemize}
 We also clarify the similarities and differences between the Newtonian and GR approaches.

\section{NEWTONIAN FRAMES VERSUS GR GAUGES}

In this section, we first define the Eulerian and Lagrangian frames in Newtonian theory and the corresponding comoving-synchronous gauge in GR. Then we consider a transformation to a gauge in GR that corresponds to an Eulerian frame. The issue of residual gauge freedom in the comoving-synchronous gauge is also discussed. 

Throughout we assume that the matter is in the form of irrotational dust, representing CDM. This is a reasonable assumption on cosmological scales. We use conformal time $\tau$ in a Friedmann-Lemaitre-Robertson-Walker background spacetime, with scale factor $a(\tau)$ and $\cH(\tau)=a'(\tau)/a(\tau)$.

\subsection{LAGRANGIAN FRAME AND COMOVING-SYNCHRONOUS GAUGE}

\noindent{\bf Newtonian theory}\\
 
In Newtonian theory, the equations that govern the dynamics of irrotational dust in an Eulerian  frame, $(\tau,x^i)$, on an expanding background are
\begin{eqnarray}
\frac{d\delta_{ N}}{d \tau} + (1+\delta_{ N}) 
\nabla^i v_{ N  i} = 0 &\quad& \mbox{(continuity)}, \\
\frac{d v^i_{ N}}{d \tau}+ {\cal H} v^i_{ N} 
+ \nabla^i \phi = 0 &\quad& \mbox{(Euler)}, \\
\nabla^2 \phi = 4 \pi G \rho a^2 \delta_{ N}
&\quad& \mbox{(Poisson)},
\end{eqnarray}
where $d/d \tau={\partial}/{\partial \tau} + v_N^i \nabla_i$ is the convective derivative and the velocity is derived from a potential:
\begin{equation}
\epsilon^{ijk} \partial_i v_{N  j} =0~~\Rightarrow~~ v_N^i=\partial^iv_N\,.
\end{equation}
In terms of the deformation tensor,
\be
\vartheta^{ij}_{N } = \partial^i v^j_{N }={1\over3}\vartheta_N\delta^{ij}+\sigma^{ij}_N,
\ee 
where $\vartheta$ is the perturbation to the expansion and $\sigma^i_j$ is the shear, 
the continuity and Euler equations can be written as 
\begin{eqnarray}
\frac{d \delta_N}{d \tau} + (1+ \delta_N) \vartheta_N &=& 0\,,\\
\frac{d \vartheta_N}{d \tau} 
+{\cal H} \vartheta_N + \vartheta^i_{N  j} \vartheta^{j}_{N  i}+ 4 \pi G \rho \delta_N &=& 0\,,
\end{eqnarray}
where we used the Poisson equation. 

In the Lagrangian picture, the dynamics is described by comoving coordinates  $q^{i}$, specified at an initial time, so that each fluid element maintains the same coordinates during evolution. Eulerian coordinates $x^i$ are related to the Lagrangian $q^i$ by 
\begin{equation}
\label{E_to_L-trasformation}
x^i (\bq, \tau) = q^i + \Psi^i (\bq, \tau)\,,
\end{equation}
where the displacement field $\Psi^i$ vanishes at the initial time, and determines the velocity in the Eulerian reference frame:
\begin{equation}\label{psi}
v_N^i \equiv v_E^i = \frac{\partial \Psi^i}{\partial \tau}. 
\end{equation}

The matter density is a scalar quantity that can be written as a function of either Eulerian coordinates or Lagrangian coordinates. These functions will have different functional dependence on the corresponding coordinates, but the density is the same physical quantity at a given point. Hence we can write the density contrast at any fixed point as
\begin{equation}
\delta_N = \delta_L(\bq,\tau) = \delta_E(\bx(\bq,\tau),\tau) \,.
\end{equation}

\noindent{\bf GR Cosmology}\\

It is possible to define a Lagrangian frame uniquely for irrotational dust in GR \footnote{It is possible to choose an irrotational growing mode for dust at first order, but at second order dust does not remain irrotational. Nonetheless we can still impose the comoving-orthogonal gauge conditions at second order with respect to the irrotational part of the velocity field.}. This is given by the comoving-synchronous gauge, with comoving  coordinates and  constant time hypersurfaces $\tau=\,$const orthogonal to the fluid worldlines. 
The line element is 
\begin{equation}\label{csm}
ds^2 = a(\tau)^2 \big[- d \tau^2 + \gamma_{i j} (\tau,q^k)dq^{i} dq^{j}\big]= a(\tau)^2 \big[- d \tau^2 + \exp \big(2\zeta(\tau,q^k) \big)\,\hat\gamma_{i j} (\tau,q^l)dq^{i} dq^{j}\big],
\end{equation}
where $\zeta$ is the comoving curvature perturbation and $|{\hat\gamma}|=1$.

The fluid four-velocity is given by $u_{\mu} = -a\delta_\mu^0$, which coincides with the unit normal to $\tau=\,$const.  
The deformation tensor is $\vartheta^\mu_\nu\equiv a\nabla_\nu u^\mu - \cH( \delta^\mu_\nu +u^\mu u_\nu)$, which in the comoving-synchronous gauge is purely spatial and is given by the extrinsic curvature~\cite{Wald:1984rg}
\begin{equation}
\vartheta^{i}_{j} 
= \frac{1}{2} \gamma^{k i} \gamma'_{k j},
\end{equation}
where a prime denotes  $\partial/\partial\tau$. 
The Einstein equations in this gauge give
\begin{eqnarray}
\vartheta^2 - \vartheta^{ij} \vartheta_{ij}
+4 {\cal H} \vartheta +  R^{(3)} =  16 \pi G a^2 \rho \delta,
\quad \mbox{(energy constraint)},\label{encon} \\
{ D}^{j} \vartheta_{ij} = 
\partial_{i} \vartheta, \quad \mbox{(momentum constraint)},\label{momcon}
\end{eqnarray}
while the evolution of the deformation tensor is given by
\begin{equation}
\vartheta^{i \; '}_{j}
+ 2 {\cal H} \vartheta^{i}_{j}
+\vartheta \vartheta^{i}_{j} 
+\frac{1}{4} \left(
\vartheta^{kl} \vartheta_{kl} - \vartheta^2 
  \right) \delta^i_j 
  +R^{(3)i}_{j} - \frac{1}{4}
  R^{(3)} \delta^{i}_{j}=0 
\label{evoln} \,,  
\end{equation}
where $R^{(3)}_{ij}$ is the Ricci tensor of $\gamma_{ij}$ and  ${D}_{i}$ is the covariant derivative defined by $\gamma_{ij}$. By combining the trace of (\ref{evoln}) and the energy constraint (\ref{encon}), we obtain the Raychaudhuri equation
\begin{equation}
\label{EulerEq}
\vartheta' +  {\cal H} \vartheta + 
\vartheta^{ij} \vartheta_{ij} + 4 \pi G a^2 \rho \delta=0. 
\end{equation}
The continuity equation is
\be
\label{ContEq}
\delta' + (1+\delta) \vartheta  = 0.
\ee 
The two equations (\ref{EulerEq}) and (\ref{ContEq}) are formally equivalent to the Newtonian equations in Lagrangian form, if we replace the comoving-synchronous time derivative in GR, $\partial/\partial\tau$, with the convective derivative in Newtonian theory, $d/d\tau$.

\subsection{EULERIAN FRAME AND TOTAL MATTER GAUGE}

In GR, any frame can legitimately be called Eulerian, while there is a unique Lagrangian frame for irrotational dust, i.e. the comoving-synchronous gauge as  discussed above (and see \cite{Ehlers:1993gf}). 
However, it is possible to define a gauge where the density contrast obeys the same evolution equations as the Newtonian theory in the Eulerian frame 
up to the second order \cite{Biern:2014zja, Rampf:2013dxa, Bruni:2013qta}.
We follow \cite{Liddle:2000cg, Malik:2008im} in calling this gauge the total matter gauge. It is also called the velocity orthogonal isotropic gauge \cite{Kodama:1985bj}  and the temporal comoving spatial $C$-gauge \cite{Yoo:2014vta}. 

In Newtonian theory the transformation between the Eulerian and Lagrangian frames is given by a spatial change of coordinates, \eqref{E_to_L-trasformation}.
We can apply the same spatial transformation to the comoving-synchronous gauge in GR, and this induces a velocity perturbation in the total matter gauge through
 \eqref{psi},  with $\partial_i v_T=\partial \Psi_i/\partial \tau$. Since  \eqref{E_to_L-trasformation} represents a purely spatial gauge transformation in GR, the two gauges share the same time-slicing, i.e.\ the  constant time hypersurfaces $\tau=\,$const are the same in the two gauges. 

There are in general two scalar degrees of freedom in the metric at each order \cite{Malik:2008im,Bruni:1996im,Matarrese:1997ay}. 
In the comoving-synchronous gauge these are both part of the spatial metric, \eqref{csm}, which (ignoring vector and tensor modes) can be written as
\begin{equation}
\gamma_{Cij} = (1+ 2 \psi_C) \delta_{ij} +  2 E_{C,ij}.
 \end{equation}
The gauge transformation \eqref{E_to_L-trasformation} induces a shift ($g_{0i}$) in the metric, and can be used to make the spatial metric conformally flat, so that the total matter gauge is completely fixed by
 \begin{eqnarray}
\psi_T=\psi_C,~~B_T = v_T,~~
E_T= E_C- \int d \tau' v_T(\bx, \tau') = 0\,, 
\end{eqnarray}
and the line element in the total matter gauge at first order becomes\footnote{In this work, we define the total matter gauge at nonlinear order as the gauge that shares the same constant time hypersurfaces as the comoving-synchronous gauge (i.e., normal to the comoving worldlines) but with spatial coordinates such that $E_T=0$ at all perturbation orders.}
\begin{equation}
ds^2 = a^2 \big[- d \tau^2  - 2 \p_i B_T dx^i d \tau 
+ (1+ 2 \psi_T) \delta_{ij}dx^idx^j   \big] \,.
\end{equation}
The density contrast and the trace of the deformation tensor (representing the perturbation in the expansion) do not change
under a purely spatial gauge transformation at first order:
\be
\delta_T=\delta_C, ~~~ \vartheta_T=\vartheta_C.
\ee
Note however that in the comoving-synchronous gauge we identify $\vartheta_C=\nabla^2E_C'$, while the total matter gauge is fixed by setting $E_T=0$ and instead we find $\vartheta_T=\nabla^2 v_T$.

The total matter gauge fixes the gauge completely but there is a residual gauge freedom in the spatial gauge transformation to the comoving-synchronous gauge,
\begin{equation}
 \label{EC}
E_C(\bx,\tau)=\int^\tau d \tilde\tau v_T(\bx, \tilde\tau) = \int^{\tau}_{\tau_{\rm in}} d \tilde\tau v_T(\bx, \tilde\tau) + \varepsilon(\bx).
\end{equation}
This can be fixed by choosing the spatial coordinates $q^i$ in the comoving-synchronous gauge to coincide with the Eulerian spatial coordinates $x^i$ 
(i.e., the spatial coordinates in the total matter gauge)
on an initial hypersurface $\tau_{\rm in}=$ const., so that 
\be
\varepsilon(\bx)=0.
\ee 
In other words, by an appropriate choice of initial conditions, this gauge mode in the comoving-synchronous gauge can be removed\footnote{
To quote Bertschinger \cite{Bertschinger:1993xt}: ``The presence of these extraneous solutions (called gauge modes) has created a great deal of confusion in the past, which might have been avoided had more cosmologists read the paper of Lifshitz \cite{Lifshitz:1945du}".
}.
This is equivalent to the usual Newtonian assumption that the Eulerian and Lagrangian coordinates coincide at an initial time, and hence the displacement field is zero initially, $\Psi^i(\bq,\tau_{\rm in})=0$.
In particular, to study the growing mode of primordial density perturbations (the regular mode at $\tau_{\rm in}\to0$), we will require $\varepsilon(\bx)=0$ in the limit $\tau_{\rm in}\to0$. 

Just as in Newtonian gravity, the density perturbation is a 3-scalar so remains invariant at a fixed physical point on the constant-time hyper-surface under the spatial coordinate transformation \eqref{E_to_L-trasformation}
\begin{equation}
\delta_C(\bq,\tau) = \delta_T(\bx(\bq,\tau),\tau) \,.
\end{equation}
At second order this gives
\begin{eqnarray}
\delta_T (\bx, \tau) 
&=& \delta_C (\bx, \tau)  -\Psi^i(\bx, \tau) \partial_i \delta_C (\bx, \tau)  \nonumber\\
&=& \delta_C (\bx, \tau) + \big[\partial_i \delta_C(\bx, \tau)\big] 
\nabla^{-2} \partial^i \delta_C
(\tau, \bx),
\label{gauge_transL-E}
\end{eqnarray}
where in the last line we have used (\ref{psi}) and the continuity equation (\ref{ContEq}) at first order,
\begin{equation}
\label{continuity_equation}
\delta'  + \vartheta= 0 \,,
\end{equation}
to eliminate the displacement field $\Psi^i$ in favour of the density contrast.
It is clear that the resulting transformation (\ref{gauge_transL-E}) is highly nonlocal at nonlinear orders.

\subsection{NEWTONIAN AND GENERAL RELATIVISTIC CONSTRAINTS}

In  Newtonian theory, the evolution equation for $\vartheta_{ij}$ is obtained by taking the spatial derivative of the Euler equation, which contains the second spatial derivative of the Newtonian potential, the tidal tensor. There is no evolution equation for the tidal tensor in  Newtonian theory and it is thus impossible to obtain a complete system of evolution equations. This is due to the underlying elliptic nature of the system, i.e. it embodies action-at-a-distance. The system is closed by a constraint, i.e. the Poisson equation. Various approximations close the evolution system using some suitable assumptions. For instance, in  the Zel'dovich approximation \cite{Zeldovich:1969sb} the evolution equation for the tidal tensor becomes redundant, and the continuity and Raychaudhuri equations, together with the evolution equation for the shear,  form a closed system of differential equations (see e.g. \cite{Bruni:2002xk}). (For a related discussion on obtaining the Newtonian equation from the GR one, see  \cite{Kofman:1994pz}.)

In GR, a covariant approach can be used, via a 1+3 split based on the fluid four-velocity $u^\mu$   \cite{Ellis:1971pg, Ellis_Maartens_MacCallum_book}. This leads to a system of evolution and constraint equations for the kinematical variables and the electric and magnetic parts of the Weyl tensor, $E_{\mu\nu}$ and $H_{\mu\nu}$. The whole set of equations is hyperbolic, i.e. it embodies causality. In GR the evolution equations for $E_{\mu\nu}$ and $H_{\mu\nu}$ close the system of evolution equations. 

Another important difference lies in the constraint equations. In the Newtonian case, the only constraint is the Poisson equation, which provides a linear relation between overdensity and potential. In GR, there are two constraints, the energy constraint (\ref{encon}) and the momentum constraint (\ref{momcon}). At first order in perturbations, the two constraints in GR combine to give the relativistic Poisson equation,
\be
\nabla^2\Phi=4\pi Ga^2\rho\delta_C,
\ee
where $\Phi$ is the metric perturbation in longitudinal gauge (see e.g. \cite{Bardeen:1980kt,Kodama:1985bj} for the gauge-invariant version).
  
However,  a difference arises at higher orders \cite{Bartolo:2005xa, Verde:2009hy, Bruni:2013qta, Bruni:2014xma}. At higher orders, the density field obeys second order differential equations that are sourced by the product of the lower order quantities. The time evolution of the homogeneous solution is given by the linear growth function $D_+(\tau)$ and its momentum dependence is determined by the constraint equation. The particular solution on the other hand describes the nonlinear evolution of the initial density contrast. The nonlinear density contrast in comoving-synchronous gauge may be written as
\begin{equation}
\label{expansion_deltaL}
\delta_C (\bq, \tau) = C(\bq) D_+(\tau) + \sum_{n=2}^\infty P_n(\bq) D_{n+}(\bq, \tau),
\end{equation}
where $C(\bq) D_+(\tau)$ is the growing mode solution of the homogeneous part of the evolution equation for the density contrast at all orders and the sum represents the nonlinear evolution of the initial density contrast at any given point $\bq$ in comoving coordinates.

On large scales, the growing mode amplitude $C(\bq)$ is related to the curvature perturbation $\zeta$ by the energy constraint equation \eqref{encon} \cite{Bruni:2014xma}:
\begin{equation}\label{czeta}
C(\bq) \propto 
\exp(-2 \zeta) \Big[-4 \nabla^2 \zeta - 2 (\nabla \zeta)^2\Big].
\end{equation}
Even in the absence of primordial non-Gaussianity, the higher-order homogeneous solution is not zero. At second order, this leads to the effective non-Gaussianity of local type $f_{\rm NL}^{\rm GR}$ in \eqref{fgr} \cite{Bartolo:2005xa, Verde:2009hy, Bruni:2013qta}. By contrast, in the Newtonian approximation, the Poisson equation is a linear relation between the density contrast and Newton potential at all orders, and therefore the higher-order homogeneous solution is zero in the absence of primordial non-Gaussianity. 

\section{Is the comoving-synchronous gauge a bad gauge? }

The comoving-synchronous gauge has been described  as inappropriate to use for analysing the growth of large-scale structure at nonlinear order \cite{Yoo:2014vta, Biern:2014zja, Hwang:2014qfa}. The specific claims include: 
\begin{itemize}  
 \item comoving-synchronous gauge ``violates  mass conservation'' \cite{Yoo:2014vta};  
 \item  the density contrast in comoving-synchronous gauge cannot be used for galaxy bias \cite{Hwang:2014qfa, Yoo:2014vta}.
\end{itemize}

In this section, we show that these claims are not correct. 

\subsection{MASS CONSERVATION}

\noindent{\bf Newtonian theory}\\

In Newtonian theory, the spatial average of the overdensity in Eulerian coordinates is  \cite{Kofman:1993mx, Bouchet:1994xp}
\begin{equation}\label{nav}
\left\langle \delta_E(\bx, \tau) \right\rangle_{x} =  \frac{1}{V} \int_V \ud^3 {\tilde x}  \, \delta_E({\bf \tilde x, \tau})\,,
\end{equation}
where the average is made over a volume that goes to infinity. 
The Lagrangian spatial average is
\begin{equation}\label{Lnav}
\left\langle \delta_L(\bq, \tau) \right\rangle_{q} =  \frac{1}{V} \int_V \ud^3 {\tilde q}  \, \mathcal{J}^{-1} \, \delta_L({\bf \tilde q, \tau})\,,
\end{equation}
where we must take account of the Jacobian for Lagrangian coordinates
\begin{equation}
\label{Jacob}
\mathcal{J}=\left|{\p x^i\over \p q^j}\right|.
\end{equation}
Taking into account that $\delta_E ({\bf x}({\bf q}, \tau),\tau)=\delta_L ({\bf q}, \tau)$, we find the relation between Eulerian and Lagrangian averages:
\begin{equation}
\label{everage-Newt}
\left\langle \delta_E ({\bf x},\tau) \right\rangle_x =\left\langle  \delta_L ({\bf q},\tau) \right\rangle_q \,.
\end{equation}

~\\
\noindent{\bf GR cosmology}\\

In GR, we use $T$- and $C$-gauges to correspond to the Eulerian and Lagrangian frames, as discussed above.
The Eulerian average 
is written using the covariant volume element:
 \begin{equation}
 \label{Tav}
\left\langle \delta_T ({\bf x},\tau) \right\rangle_x =  \frac{1}{V} \int_V \ud^3 \tilde \bx \, \left|\gamma_T\right|^{1/2}  \delta_T({\bf \tilde x},\tau),
\end{equation}
where $\gamma_T$ is the metric induced on $V$ in the $T$-gauge, while in the Lagrangian $C$-gauge we have
 \begin{equation}
 \label{averageGR}
\left\langle \delta_C ({\bf q},\tau) \right\rangle_q=
\frac{1}{V} \int_V \ud^3 \tilde \bq \, \left|\gamma_C\right|^{1/2}  \delta_C ({\bf \tilde q},\tau) \,.
\end{equation}
Note that at first order,
\begin{equation}
\frac{\left|\gamma_C\right|^{1/2}}{\left|\gamma_T\right|^{1/2}} = 1+\nabla^2 E_C  \,,
\end{equation}
and hence using (\ref{EC}) and (\ref{psi}) we can identify this with the Jacobian (\ref{Jacob})
\begin{equation}
\mathcal{J} = 1+\partial_i\Psi^i \,.
\end{equation}
Actually, $\mathcal{J}=(|\gamma_C|/|\gamma_T|)^{1/2}$ is a general result in tensor calculus for the coordinate transformation of the determinant of a metric $\gamma_{ij}$.
In the transformation to the comoving-synchronous gauge, the volume element is invariant
\begin{equation}
\ud^3 \tilde \bx \, \left|\gamma_T\right|^{1/2} = \ud^3 \tilde \bq \, \left|\gamma_C\right|^{1/2} \,,
\end{equation}
since the change of coordinates is purely spatial.
Thus we have, as in Newtonian theory,
\begin{equation}
 \label{LtoEframe}
\langle\delta_T({\bf x},\tau)\rangle_x
 = \left\langle \delta_C({\bf q},\tau) \right\rangle_q.
 \end{equation}
The spatial average of the density contrast in the total matter gauge must equal the spatial average in the comoving-synchronous gauge since we are evaluating the same physical quantity on the same physical hypersurface, just using different spatial coordinates.

The solution up to second order in the total matter gauge (restricting for simplicity to the case of planar symmetry or matter-dominated era) can be written in Fourier space as
\begin{equation}
\hat\delta_T(\bk,\tau) = D_+(\tau)\hat\delta^{(1)}(\bk) 
+ D_+^2(\tau) \int \ud^3\tilde {\bf k}
\, F_{2}^{T}(\tilde {\bf k},\bk-{\bf \tilde k}) \, \hat\delta^{(1)} ({\bf \tilde k}) \, \hat\delta^{(1)} (\bk-{\bf \tilde k}) \;.
\end{equation}
Here we use the Fourier transform with respect to Eulerian coordinates
\be
  \hat{g}(\bk)=\int \frac{\ud^3{\bf x}}{(2 \pi)^3} e^{-i \bk \cdot \bx}  g(\bx) \,.
\ee
The Eulerian kernel vanishes in the large-scale limit
\bea
\lim_{k_2\to0}
F_{2}^{T}(\bk_1,\bk_2-\bk_1) & 
=
& \frac{1}{7} \left[3-5 \frac{(\bk_1\cdot \bk_2)^2}{k_1^2 k_2^2}\right] \frac{k_2^2}{k_1^2},
\eea
and hence the spatial average vanishes in the large-scale limit.

In  \cite{Yoo:2014vta} it is argued that the kernel for the comoving-synchronous gauge density contrast does not vanish in the large-scale limit:
\bea
\lim_{k_2\to0}
F_{2}^{C}(\bk_1,\bk_2-\bk_1)& 
=
& 1+  \frac{2}{7}  \left[ \frac{(\bk_1\cdot \bk_2)^2}{k_1^2 k_2^2}-1\right] \frac{k_2^2}{k_1^2}
\label{ysc} 
\eea
and thus the spatial average does not vanish at second order, violating mass conservation.

This misunderstanding comes from taking the Fourier transform in Eulerian coordinates, corresponding to the spatial average $\left\langle  \delta_C ({\bf x},\tau) \right\rangle_x$.
Using the second-order gauge transformation \eqref{gauge_transL-E}, we note that
 \begin{equation}
 \langle\delta_T({\bf x},\tau)\rangle_x
 = 
 \left\langle  \delta_C ({\bf x},\tau)  - \Psi^i({\bf x},\tau)\, \p_{i} \delta_C ({\bf x},\tau)  \right\rangle_x
= \left\langle \delta_C({\bf q},\tau) \right\rangle_q
\,.
\end{equation}
Thus using  (\ref{psi}) and (\ref{continuity_equation}),  we find that the apparent difference in the spatial average of $\delta_T$ and $\delta_C$ in Eulerian coordinates is due to a dipole term, which can be written in Fourier space as
\begin{eqnarray}
\label{dipole}
&&\widehat{\left[ \Psi^i \, \p_{i} \delta_C \right]}({\bf k}, \tau) 
= D_{+}^2(\tau)\int{\ud^3\tilde {\bf k}_1 \over(2\pi)^3} \left[{\tilde{\bf k}_1\cdot \tilde {\bf k}_2 \over 2 \tilde k_1 \tilde k_2}\left({\tilde k_1 \over \tilde k_2}+{\tilde k_2\over  \tilde k_1}\right)\right] \hat\delta_C ({\bf \tilde k}_1) \, \hat\delta_C ({\bf \tilde k}_2) \;,
\end{eqnarray}
where ${\bf \tilde k}_2={\bf k}-{\bf \tilde k}_1$. Clearly this does not vanish in the large-scale limit, $k\to0$.
Because of this dipole term,  \cite{Yoo:2014vta} concludes that the matter density fluctuation in the comoving-synchronous gauge violates mass conservation at second order. 

The problem with naively using \eqref{ysc} in \eqref{Tav} is that this is the Eulerian average relation, instead of the correct Lagrangian average \eqref{averageGR}. The result \eqref{dipole} shows clearly that the dipole term is {\em not} missing in comoving-synchronous gauge --
 we can recover the {\em same} kernel that we get in the total matter gauge (see, for example, the appendix of \cite{Bruni:2013qta}). 
This proves that,  in both the total matter and the comoving-synchronous gauges, the kernel $F_2$ vanishes as $k^2$ in the large-scale limit. Therefore we recover the correct result in either gauge without any violation of mass conservation (which is of course respected in GR in any gauge).

\subsection{LAGRANGIAN GALAXY BIAS}

In Newtonian theory there are two common prescriptions for describing the distribution of galaxies (assumed to reside in collapsed dark matter halos) as biased tracers of the underlying matter density. 
An Eulerian bias model assumes that the local overdensity of galaxies traces the local matter overdensity at a given Eulerian point ${\bf x}$, while in a Lagrangian bias model it is the initial overdensity in the Lagrangian coordinates that determines the galaxy distribution.

As we have seen, the total matter gauge in GR corresponds to an Eulerian frame. 
On the other hand, Lagrangian bias models follow the collapse of initial overdensities in the matter field at a fixed point ${\bf q}$ in Lagrangian coordinates, i.e., comoving with the matter distribution (which moves through Eulerian space). Therefore the comoving-synchronous gauge provides the simplest GR gauge in which to describe Lagrangian bias. 
In the 
Press-Schecter theory \cite{Press:1973iz}, for example,
halos of a given mass, $M$, collapse when the linearly growing local density contrast $\delta^{\rm lin}(\bq)$ (smoothed on the corresponding mass scale) reaches a critical value. 
In particular, spherical collapse has an exact GR solution in the comoving-synchronous gauge \cite{Wands:2009ex}.

The local galaxy number density in comoving-synchronous coordinates, $n_{g\,C}({\bf q})$, is thus a function of the initial density field, corresponding to a local Lagrangian bias model~\cite{Kofman:1993mx,Bouchet:1994xp, Catelan:1997qw, Matsubara:2011ck}
\begin{equation}
\label{LagrangianBias1}
\delta_{g\,C} (\bq) = b_{1}^L \delta^{\rm lin} (\bq) + b_{2}^L\big[ \delta^{\rm lin} (\bq) \big]^2 +\ldots\;.
\end{equation}
Since the initial displacement between Lagrangian and Eulerian frames is zero, at sufficiently early times where we can also neglect nonlinear evolution, the density field is the same in both comoving-synchronous and total matter gauges, 
where $\bx=\bq$ initially.
However as the density contrast evolves, a local Lagrangian bias model in general predicts a nonlocal Eulerian bias \cite{Catelan:2000vn,Matsubara:2011ck}. 

Note that the Lagrangian bias model describes the number density $n_g$ with respect to an initial (Lagrangian) coordinate volume and thus $n_g$ is not a 3-scalar in GR, but rather a density of weight one.
Transforming the number density from Lagrangian to Eulerian coordinates, we must include the Jacobian:
\begin{equation}
n_{g\,T}({\bf x}) =  n_{g\,C}({\bf q}({\bf x})) {\cal J}^{-1},
\end{equation}
and thus in terms of the galaxy overdensity we have
\begin{equation}
1+ \delta_{g\,T}({\bf x}) = [1+\delta_T({\bf x})][1+ \delta_{g\,C}({\bf q}({\bf x}))].
\end{equation}
Transforming the Lagrangian galaxy density contrast $\delta_{g\,C}$ from Lagrangian ${\bf q}$  to Eulerian ${\bf x}$ using \eqref{gauge_transL-E}, then introduces nonlocal terms at second and higher order in the Eulerian galaxy density contrast $\delta_{g\,T}({\bf x},\tau)$, as shown for example in \cite{Catelan:1997qw,Catelan:2000vn,Matsubara:2011ck}:
\begin{equation}
\label{EulerianBias1}
\delta_{g\,T} = b_{1}^E \delta_T + b_{2}^E \big[ \delta_T \big]^2 + b_s s^2 + \ldots,
\end{equation}
where 
$s^2$ describes second-order tidal terms.

The local Lagrangian bias model has been shown to give a better fit to the distribution of collapsed halos in N-body simulations than a local Eulerian bias model 
\cite{Baldauf:2011bh,Chan:2012jj}.
For complete generality, the Lagrangian bias model may also include nonlocal terms due to tidal effects  \cite{Sheth:2012fc}.

The initial density contrast (in Lagrangian or Eulerian frames) is dominated by the linearly growing mode\footnote{Here the ``linearly growing mode'' indicates that the time evolution is given by the linear growth function $D_+(\tau)$.}:
\begin{equation}
\delta^{\rm lin}(\bq,\tau) = C(\bq) D_+(\tau) .
\end{equation}
In GR, unlike in Newtonian gravity, the growing mode amplitude $C(\bq)$ is nonlinearly related to the primordial curvature perturbation $\zeta$ 
by \eqref{czeta}. Thus even if $\zeta$ is Gaussian, the initial density field is non-Gaussian. There has been significant interest in the effect of non-Gaussianity of the initial density field in GR on galaxy bias \cite{Bartolo:2005xa, Verde:2009hy, Bruni:2013qta, Bruni:2014xma}. It is beyond the scope of this paper to discuss in details this effect and we leave it to future work.

\section{Conclusions}

Recently, several papers have extended the calculation of the observed galaxy number overdensity on cosmological scales up to second order in relativistic perturbation theory \cite{Bertacca:2014dra, Bertacca:2014wga, Yoo:2014sfa, DiDio:2014lka, Bertacca:2014hwa}.
These results  include all GR effects that arise from observing on the past lightcone and will be important for accurate cosmological parameter estimation, including non-Gaussianity. In order to pursue such applications, it is important to carefully choose the gauge in which to relate the fluctuations of galaxy number density to the underlying matter density fluctuation. 

For an irrotational dust flow in $\Lambda$CDM, assuming  that the galaxy velocity is equal to the CDM velocity on large scales,  \cite{Bertacca:2014dra} claimed  that a suitable gauge to define the local bias up to second order is  the comoving-synchronous  gauge. This choice was criticised in  \cite{Yoo:2014vta}, with the claim that the comoving-synchronous gauge ``does not properly represent the matter density fluctuation with the mean at the local proper time, violating  mass conservation", and the further claim that the correct  gauge to choose is  the total matter gauge. 

We have shown that the comoving-synchronous  gauge:
\begin{itemize}
 \item  is the gauge in GR that best matches the concept of Lagrangian frame in Newtonian theory, 
 \item  does {\em not} violate any  mass conservation at second order, and 
 \item is appropriate for defining local Lagrangian galaxy bias up to second order.
 \end{itemize}

As part of our argument, we have highlighted the differences between the Newtonian approximation and the correct General Relativistic analysis.

\section*{Acknowledgments}

The authors thank  Juan Carlos Hidalgo, Cornelius Rampf,  Eleonora Villa and Jaiyul Yoo for useful comments and discussions.
DB and RM  were supported by the South African Square Kilometre Array Project. DB acknowledges financial support from the Deutsche Forschungsgemeinschaft through the Transregio 33, {\em The Dark Universe}. RM is supported by the South African National Research Foundation. MB, KK, RM, and DW are supported by the UK Science \& Technology Facilities Council grants ST/K00090X/1 and ST/L005573/1. 
MS is supported by JSPS Grant-in-Aid for Scientific Research (A) No.~21244033.
NB and SM acknowledge partial financial support by the ASI/INAF Agreement 2014-024-R.0 for the Planck LFI Activity of Phase E2. 
MB, KK, MS and DW benefited from discussions during the workshop {\em Relativistic Cosmology} (YITP-T-14-04) at the Yukawa Institute for Theoretical Physics, Kyoto University. 

~\\{\bf Note:} While this paper was being prepared, some similar points were made in \cite{Rampf:2014mga}.

\end{document}